\documentclass[aps,prl,twocolumn,superscriptaddress,pdftex]{revtex4-2}

\usepackage[pdftex]{graphicx}
\usepackage{physics}
\usepackage{amsmath}
\usepackage{amssymb}
\usepackage[pdftex]{hyperref}
\usepackage{xcolor}
\hypersetup{
    colorlinks=true,
    citecolor=blue,
    linkcolor=red,
    urlcolor=blue
}

\begin{document}


\title{Scalable implementation of $(d+1)$ mutually unbiased bases\\for $d$-dimensional quantum key distribution}


\author{Takuya Ikuta}
\email[]{takuya.ikuta.xu@hco.ntt.co.jp}
\affiliation{NTT Basic Research Laboratories, NTT Corporation, 3-1, Morinosato Wakamiya, Atsugi, Kanagawa, 243-0198 Japan}
\affiliation{Graduate School of Engineering, Osaka University, 2-1, Yamadaoka, Suita, Osaka, 565-0871 Japan}

\author{Seiseki Akibue}
\email[]{seiseki.akibue.rb@hco.ntt.co.jp}
\affiliation{NTT Communication Science Laboratories, NTT Corporation, 3-1, Morinosato Wakamiya, Atsugi, Kanagawa, 243-0198 Japan}

\author{Yuya Yonezu}
\author{Toshimori Honjo}
\author{Hiroki Takesue}
\affiliation{NTT Basic Research Laboratories, NTT Corporation, 3-1, Morinosato Wakamiya, Atsugi, Kanagawa, 243-0198 Japan}

\author{Kyo Inoue}
\affiliation{Graduate School of Engineering, Osaka University, 2-1, Yamadaoka, Suita, Osaka, 565-0871 Japan}


\date{\today}

\begin{abstract}
A high-dimensional quantum key distribution (QKD) can improve error rate tolerance
and the secret key rate.
Many $d$-dimensional QKDs have used two mutually unbiased bases (MUBs),
while $(d+1)$ MUBs enable a more robust QKD, especially against correlated errors.
However,
a scalable implementation has not been achieved because
the setups have required $d$ devices even for two MUBs
or a flexible convertor for a specific optical mode.
Here, we propose a scalable and general implementation of $(d+1)$ MUBs
using $\log_p d$ interferometers in prime power dimensions $d=p^N$.
We implemented the setup for time-bin states
and observed an average error rate of 3.8\% for phase bases,
which is lower than the 23.17\%
required for a secure QKD against coherent attack in $d=4$.
\end{abstract}


\maketitle

Quantum key distribution (QKD) is a technique to share a secret key,
whose security is guaranteed by quantum mechanics.
Ever since the first proposal, called the BB84 protocol \cite{Bennett1984a},
many different types of protocols have been proposed and demonstrated \cite{Ekert1991,Brus1998,Enzer2002,Inoue2002,Takesue2007,Schmitt-Manderbach2007,Lo2012,Sasaki2014,Takesue2015b,Yin2016,Liao2017b,Boaron2018}.
One of the key ingredients for QKD is mutually unbiased bases (MUBs) \cite{Wootters1989,Durt2005a,DURT2010}.
If two $d$-dimensional states
$\forall \ket{\psi} \in \mathcal{B}_0$,
$\forall \ket{\phi} \in \mathcal{B}_1$
for orthonormal bases $\mathcal{B}_0, \mathcal{B}_1$ satisfy
$\left|\braket{\psi}{\phi}\right|^2=1/d$,
the two bases are mutually unbiased.
Typical examples of MUBs are sets of eigenstates of the Pauli operators,
$\sigma_x, \sigma_y$, and $\sigma_z$.
Generally,
at most $(d+1)$ bases can be mutually unbiased \cite{Wootters1989},
where any pair of bases satisfies the above condition.
The BB84 protocol employs two two-dimensional MUBs (e.g., X and Z bases),
while the six-state protocol employs three two-dimensional MUBs (X, Y and Z bases) \cite{Brus1998}.
In the case of standard one-way error correction,
the error rate threshold to distill a secure key is 11.0\% for the BB84 protocol
while the six-state protocol enhances the threshold to 12.6\%
for a depolarizing channel.
In particular,
the six-state protocol has a larger error tolerance
when bit- and phase-flip errors are correlated \cite{Koashi2009};
thus, using a larger set of MUBs improves the robustness and secret key rate of QKD.

High-dimensional quantum states open another way to improve the secret key rate \cite{Cerf2002,Sheridan2010,Ferenczi2012,Wang2021},
where the information amount per photon increases with dimensions.
A two-basis protocol, which is a $d$-dimensional extension of the BB84 protocol,
uses two MUBs in $d$ dimensions to ensure security.
Similar to the two-dimensional QKDs,
the secret key rate of a high-dimensional QKD can be further improved by using $(d+1)$-MUBs,
especially against correlated noises (see \cite{SupplementalMaterial} for an example of correlated noises).

The two-basis protocol has been performed using several optical modes \cite{Etcheverry2013,Mirhosseini2015,Sit2016a,Islam2017,Islam2017a,Ding2017,Canas2017,Vagniluca2020,Hu2021,Wang2021a}.
Among these optical modes,
a time-bin state is a promising candidate for QKD
because of its high robustness against disturbances during fiber transmission \cite{Inagaki2013,Ikuta2018}.
Unfortunately, increasing dimensions for time-bin states consumes many time slots, 
and the improvement of the key rate per unit time is limited if we can use ideal devices.
However, dead time due to practical single-photon detectors and slow electrical devices
limits the detection count rates in a short distance.
In such a situation,
the large amount of information per photon in a high-dimensional time-bin state
makes it possible to directly improve the secret key rate
because we can use many short pulses within the dead time thanks to fast optical devices \cite{Islam2017a}.
On the other hand,
a robust $(d+1)$-basis protocol has also been implemented using orbital angular momentum (OAM)
thanks to the flexibility of spatial light modulators (SLMs) \cite{Mafu2013}.

Although the advantages of high-dimensional QKD have been demonstrated,
important problems still remain in scaling up the dimensions.
One problem is the number of devices required for the measurement.
For example, the Fourier basis $\mathcal{F}$ used in \cite{Islam2017a} is given by
\begin{equation}
    \mathcal{F} = \left\{\ket{f_n} = \frac{1}{\sqrt{d}} \sum_{m} e^{\frac{2\pi i mn}{d}}\ket{m} \middle| n\in \{0,\cdots, d-1\} \right\} , \label{eq:fourie}
\end{equation}
where $\left\{\ket{m} \middle| m\in \{0,\cdots, d-1\} \right\}$ is the Z basis selected as the time-bin basis
representing a photon in a specific time slot.
To measure a time-bin state in $\mathcal{F}$,
$(d-1)$ delay Mach-Zehnder interferometers (MZIs) and $d$ single-photon detectors are required;
thus, the number of devices increases linearly with dimensions \cite{Islam2017,Islam2017a}.
Another problem is a requirement for precise control and calibration of devices
because the phase in Eq. \eqref{eq:fourie} is $\propto 1/d$.
Some approaches mitigated these problems by using other bases for specific dimensions \cite{Vagniluca2020}
or using fewer states instead of all the states in a basis \cite{Islam2018,Islam2019}.
However, a generally scalable method has not been established to the best of our knowledge.
Furthermore, a generalization to $(d+1)$ MUBs is more challenging,
especially
for time-bin states because they cannot be modulated flexibly as in the case of OAM using an SLM
in the current technology,
although, in principle,
a universal unitary gate can be implemented by optical switches and many $2\times 2$ beam splitters (on order of $\mathcal{O}(d^2)$) \cite{Bussieres2006}.
In addition,
previous analyses of the $(d+1)$-basis protocol were
limited to prime dimensions \cite{Sheridan2010,Ferenczi2012,Wang2021},
which largely restricts the dimensions available for the $(d+1)$-basis protocol.

Here,
we propose an implementation of $(d+1)$ MUBs for prime power dimensions $d=p^N$,
where the number of interferometers scales logarithmically with $d$
and the phase resolution is constant regardless of $N$.
The proposed method can also be applied to general optical modes.
In addition,
the required number of detectors is constant regardless of $N$ in the case of time-bin states.
We also show that the MUBs we use can ensure security against coherent attack in prime power dimensions.

To implement a compact setup,
we use MUBs constructed by the Galois field \cite{Wootters1989,Durt2005a,DURT2010}.
Because the equation representing MUBs depends on
whether $p$ is 2 or odd prime number,
we first explain the case of $d=2^N$,
and the case of odd prime numbers is explained later.
Let $GF[2^N]$ be the Galois field of order $2^N$,
where the addition is elementwise exclusive OR in a bit representation of $e \in GF[2^N]$.
We also define a binary symmetric matrix $\mathbf{A}^{(k)}$ whose element satisfies
$2^i \odot 2^j = \oplus_{k=0}^{N-1} A_{ij}^{(k)} 2^k$,
where $\oplus$ and $\odot$ denote the addition and multiplication in $GF[2^N]$, respectively.
We select the Z basis as one basis in $(d+1)$ MUBs.
Then, all states in other $d$ MUBs (phase bases) are represented by
$\ket{\psi_n^{(r)}} = \sum_m B_{mn}^{(r)} \ket{m}$,
where $r, n \in GF[2^N]$ are the labels of the phase basis and state, respectively.
If we define the probability amplitude by the following equation,
the Z and phase bases form $(d+1)$ MUBs \cite{Wootters1989}.
\begin{equation}
    B_{mn}^{(r)} = \frac{1}{\sqrt{2^N}} \exp(\frac{\pi}{2}i \left(\sum_{j=0}^{N-1} r_j \mathbf{m}^\mathrm{T} \mathbf{A}^{(j)} \mathbf{m} + 2\mathbf{m} \cdot \mathbf{n} \right)) . \label{eq:Bmnr}
\end{equation}
Here, $\mathbf{m}$ and $\mathbf{n}$ are binary vectors representing $m$ and $n$, respectively,
and $r_j$ is $j$th element of $r$ in bit representation.
In contrast to the phase resolution of $1/2^N$ in Eq. \eqref{eq:fourie},
it is clear that the phase in Eq. \eqref{eq:Bmnr} takes only four values, $\{0, \pi/2, \pi, 3\pi/2\}$,
because $r_j, \mathbf{m}, \mathbf{n}$, and $\mathbf{A}^{(j)}$ are a binary value, vector, and matrix.
Therefore, these states can be easily generated for any large $N$.
For example, if we use the time-bin basis as the Z basis,
the phase basis state can be generated by quadrature phase shift keying (QPSK) modulation of $d$ sequential pulses.

For a compact measurement of phase bases,
we use a decomposition of the matrix $\mathbf{B}^{(r)}$ whose element is $B_{mn}^{(r)}$.
$\mathbf{B}^{(r)}$ can be decomposed as $\mathbf{D}^{(r)}\mathbf{B}^{(0)}$,
where $\mathbf{D}^{(r)}$ is a diagonal unitary and the $m$th element of $\mathbf{D}^{(r)}$ is given by
\begin{equation}
    D_{mm}^{(r)} = \exp(\frac{\pi}{2}i \left(\sum_{j=0}^{N-1} r_j \mathbf{m}^\mathrm{T} \mathbf{A}^{(j)} \mathbf{m} \right)).  \label{eq:diag_unitary}
\end{equation}
This decomposition implies that
the measurement of the $r$th phase basis can be implemented by two sequential procedures:
the first procedure is a basis selection corresponding to $\mathbf{D}^{(r)}$,
and the second procedure is a projective measurement onto $\left\{ \ket{\psi_n^{(0)}} \right\}$ corresponding to $\mathbf{B}^{(0)}$.
Since $\mathbf{D}^{(r)}$ is a diagonal unitary,
the basis change is a phase modulation on Z basis states.
If Z basis states are time-bin states,
$\mathbf{D}^{(r)}$ can be implemented by an optical phase modulator.
Regarding the projective measurement,
we can confirm that $\left\{ \ket{\psi_n^{(0)}} \right\}$ are the Hadamard basis states
because $B_{mn}^{(0)} = \frac{1}{\sqrt{2^N}} \exp(\pi i \mathbf{m} \cdot \mathbf{n} )$ is the Hadamard transform matrix.
Therefore,
$\left\{ \ket{\psi_n^{(0)}} \right\}$ can be readily represented in
$N$-qubit space equivalent to the $d$-dimensional Hilbert space.

\begin{figure}[tbp]
    \centering
    \includegraphics[width=8.6cm,pagebox=cropbox,clip]{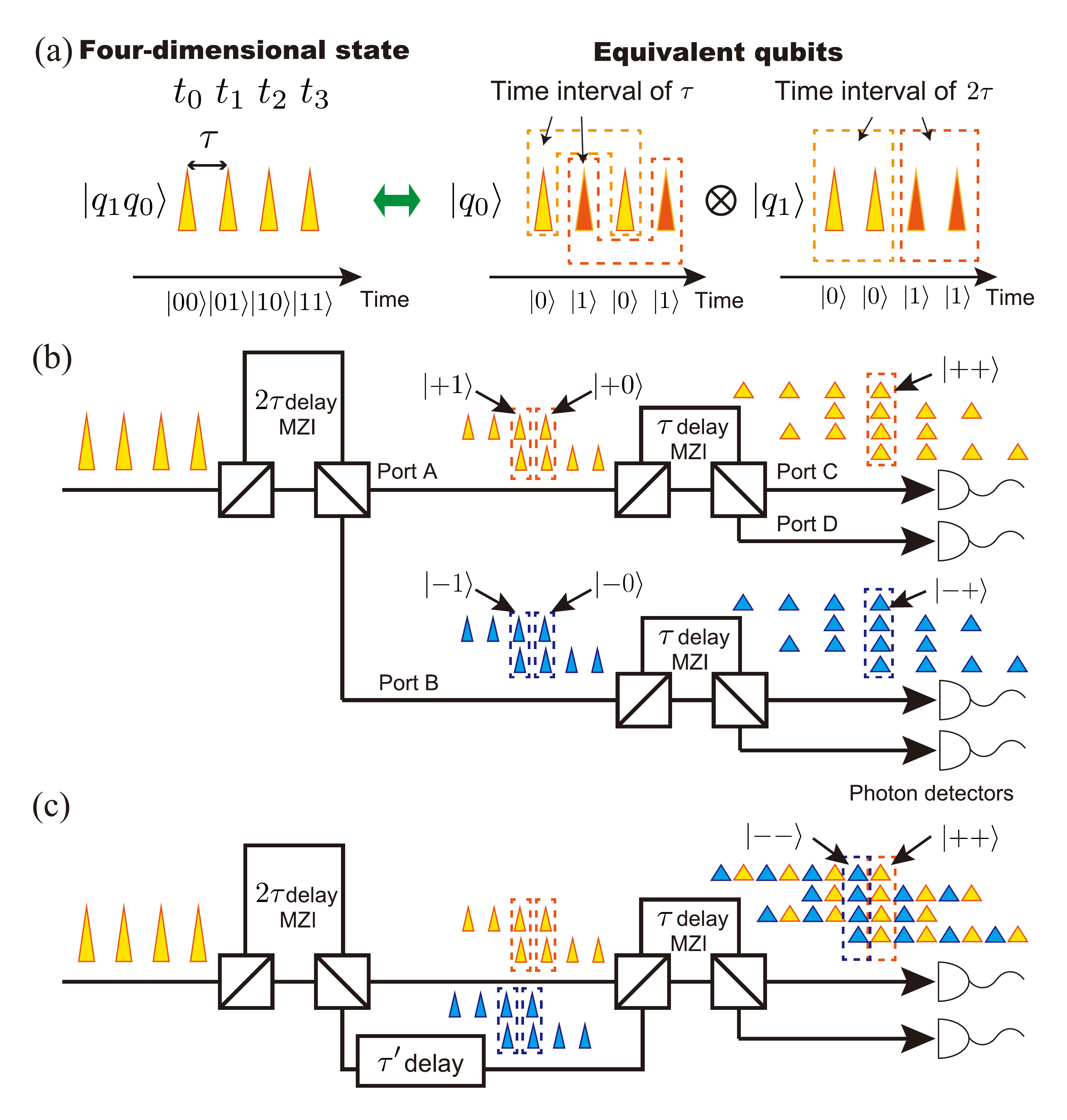}
    \caption{(a) Two blockwise qubits equivalent to a four-dimensional time-bin state.
    Measurement setup for the Hadamard basis using cascaded MZIs with (b) a tree structure
    and (c) time division multiplexing.}
    \label{fig:equivalent_qubits_and_CMZIs}
\end{figure}

For a more concrete explanation,
let us consider two qubits equivalent to a four-dimensional time-bin state (Fig. \ref{fig:equivalent_qubits_and_CMZIs}(a)).
We assign two-qubit states $\ket{00}, \ket{01}, \ket{10}, \ket{11}$ to
a photon existing in time slots $t_m$ in ascending order.
Here, these two qubits span the four-dimensional Hilbert space.
When we look at each qubit,
each qubit is a conventional time-bin qubit except that $\ket{0}$ and $\ket{1}$ correspond to blocks of several time slots.
For example,
$\ket{0}$ and $\ket{1}$ for the first qubit $q_0$ correspond to $\{t_0, t_2\}$ and $\{t_1, t_3\}$, respectively,
and the time interval between the blocks is $\tau$.
Similarly,
the second qubit $q_1$ is a time-bin qubit with a different time interval, $2\tau$.
The Hadamard transform converts $\ket{0}$ and $\ket{1}$ to $\ket{\pm}=\frac{1}{\sqrt{2}}\left(\ket{0} \pm \ket{1} \right)$ for each qubit.
Therefore, $\ket{\psi_n^{(0)}}$ is the tensor products of $\ket{\pm}$ for these qubits.
Because a projection onto $\ket{\pm}$ can be performed by a delay MZI having the same delay time as the interval of a time-bin qubit \cite{Takesue2009},
the projective measurement onto $\left\{ \ket{\psi_n^{(0)}} \right\}$ can be implemented by cascading several MZIs.

Fig. \ref{fig:equivalent_qubits_and_CMZIs}(b) shows
the measurement setup for $\left\{ \ket{\psi_n^{(0)}} \right\}$ by cascaded MZIs using a tree structure.
First, a four-dimensional time-bin state is launched into an  MZI with a delay time of $2\tau$.
By setting the relative phase between the two arms at 0,
the interferences at port A correspond to projections onto $\ket{+0}$ and $\ket{+1}$,
while those at port B correspond to projections onto $\ket{-0}$ and $\ket{-1}$.
Then, these states are launched into another MZI with a delay time of $\tau$.
As a result,
the projections onto $\ket{++}$ and $\ket{+-}$ are performed at port C and D, respectively.
Similarly,
the projections onto $\ket{-+}$ and $\ket{--}$ are performed at the other MZI with a delay time of $\tau$.
By expanding the tree structure,
we can implement projective measurements onto the tensor product states of $\ket{\pm}$ for any $2^N$ dimensions.
However,
this implementation is not scalable because it requires $(d-1)$ MZIs,
the same as the implementation of the Fourier basis measurement does \cite{Islam2017a,Islam2017}.
An important difference between these implementations is that the Hadamard basis measurement is implemented simply with the relative phase of 0,
while the Fourier basis measurement requires several different phases.
Because the same MZIs are used several times for the Hadamard basis measurement,
we can reduce the number of MZIs significantly (Fig \ref{fig:equivalent_qubits_and_CMZIs}(c)).
In this setup,
each output of the first MZI is connected to each input of the second MZI after the input timings are adjusted by an optical delay line.
Here,
the delay time $\tau'$ is chosen so that the two inputs for the second MZI do not make interferences.
By concatenating this time-division multiplexing,
all projections onto $\left\{ \ket{\psi_n^{(0)}} \right\}$ for general $2^N$ dimensional states can be implemented using $\log_2 d$ MZIs and two single-photon detectors.
Combined with the basis selection of $\mathbf{D}^{(r)}$,
$d$ phase basis measurements can be implemented by a significantly compact setup using a phase modulator and cascaded MZIs.
The remaining measurement on the time-bin basis can be performed by simply launching the photon into a single-photon detector,
and the Z and phase basis measurements can be simultaneously implemented by a beam splitter.

Note that the method using cascaded interferometers is also available for other optical modes.
The general structure is as follows.
$\mathbf{D}^{(r)}$ is implemented by a mode-dependent phase modulator.
The projective measurement onto $\left\{\ket{\psi_n^{(0)}}\right\}$ is implemented by cascading $\log_2 d$ MZIs having different mode shifts connected via delay lines or optical lines with mode shifts \cite{SupplementalMaterial}.

In the proposed method,
the discarded time slots at the outputs of the MZIs reduce the detection efficiency,
whose equivalent optical loss is $N \times 3$ dB.
It is known that, ideally,
this inefficiency of the Hadamard transform can be removed by replacing the input beam splitters in MZIs with active optical switches \cite{Bussieres2006,Dissertation2016}.
For example,
if we replace the first beam splitter of the $2\tau$-delay MZI in Fig \ref{fig:equivalent_qubits_and_CMZIs}(c),
the optical switch can transfer the pulses corresponding to $\ket{0}$ and $\ket{1}$ for the equivalent qubit $q_1$
to the long and short arms, respectively.
As a result,
the outputs from the $2\tau$-delay MZI contain the optical pulses
only for the time slots surrounded by dotted lines in Fig \ref{fig:equivalent_qubits_and_CMZIs}(c).
The same operation can be implemented for the $\tau$-delay MZI and the equivalent qubit $q_0$.
By iterating the same operation for all equivalent qubits,
we can avoid the inefficiency of the Hadamard transform for the passive implementation.
Although active devices usually introduce additional insertion losses,
such a modification is beneficial if the insertion loss per switch is less than 3 dB.

\begin{figure*}[!htbp]
    \centering
    \includegraphics[width=18cm,pagebox=cropbox,clip]{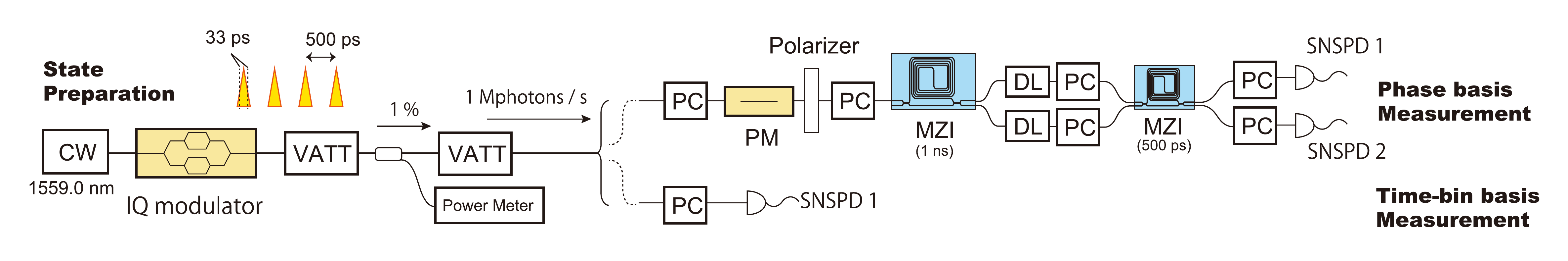}
    \caption{Experimental setup for the four-dimensional MUBs using a time-bin state.
    The dotted lines in the middle were manually selected.
    CW, continuous-wave light;
    IQ modulator, $\mathrm{LiNbO_3}$ in-phase and quadrature phase modulator;
    VATT, optical variable attenuator;
    PC, polarization controller;
    PM, $\mathrm{LiNbO_3}$ phase modulator;
    MZI, Mach-Zehnder interferometer fabricated by planar lightwave circuit technology;
    DL, optical delay line;
    SNSPD, superconducting nanowire single photon detector.}
    \label{fig:experimental_setup}
\end{figure*}

We implemented the proposed setup for four-dimensional time-bin states (see \cite{SupplementalMaterial} for more details, including calibration).
A continuous-wave light, whose wavelength was 1559.0 nm,
was modulated into a single pulse or four sequential pulses by an optical in-phase and quadrature phase (IQ) modulator (Fig. \ref{fig:experimental_setup}).
A single pulse corresponded to the time-bin basis state,
while a set of four-sequential pulses corresponded to the phase basis state represented by Eq. \eqref{eq:Bmnr}.
The pulse width, time interval, and repetition frequency of the state preparation were 33 ps, 500 ps, and 250 MHz, respectively.
The optical power was attenuated by variable optical attenuators and optical couplers
so that the average photon number became 1 M photons per second.

The prepared states were measured by two measurement setups.
The first setup was used to measure the photons in the phase bases.
The weak optical pulses were launched into a $\mathrm{LiNbO_3}$ phase modulator (PM),
by which we implemented the phase basis selection $\mathbf{D}^{(r)}$.
The modulated pulses were then launched into a stable MZI fabricated by using a planar lightwave circuit (PLC) technology \cite{Honjo2004,Takesue2005}.
The delay time was 1 ns, and the relative phase between the two arms was adjusted to be 0.
The pulses output from this MZI were launched into another MZI
after a relative delay of $\approx 250$ ps was introduced by two optical delay lines (DLs).
The delay time of the second MZI was 500 ps,
and the relative phase was also adjusted to be 0.
The outputs from these cascaded MZIs were detected by using two superconducting nanowire single-photon detectors (SNSPDs).
The detection efficiencies were set at 56\%, and the dark counts were $<$ 100 cps.
The single count rates for SNSPD1 and 2 were 52 and 47 kcps, respectively;
thus, the dark counts were negligible.
These counts were recorded by a time-interval analyzer.
The second setup was used to perform the time-bin basis measurement,
where the photons were directly detected by SNSPD1 after removing the PM and MZIs.
The single count rate was 546 kcps.
Due to the timing jitter of the detectors and electrical system,
we observed the full-width-at-half-maximum temporal width of 78 ps for the histogram of the arrival time records. 
As we described above,
this measurement can be simultaneously performed by using a beam splitter when we implement an actual QKD protocol.

\begin{figure}[btp]
    \centering
    \includegraphics[width=8.6cm,pagebox=cropbox,clip]{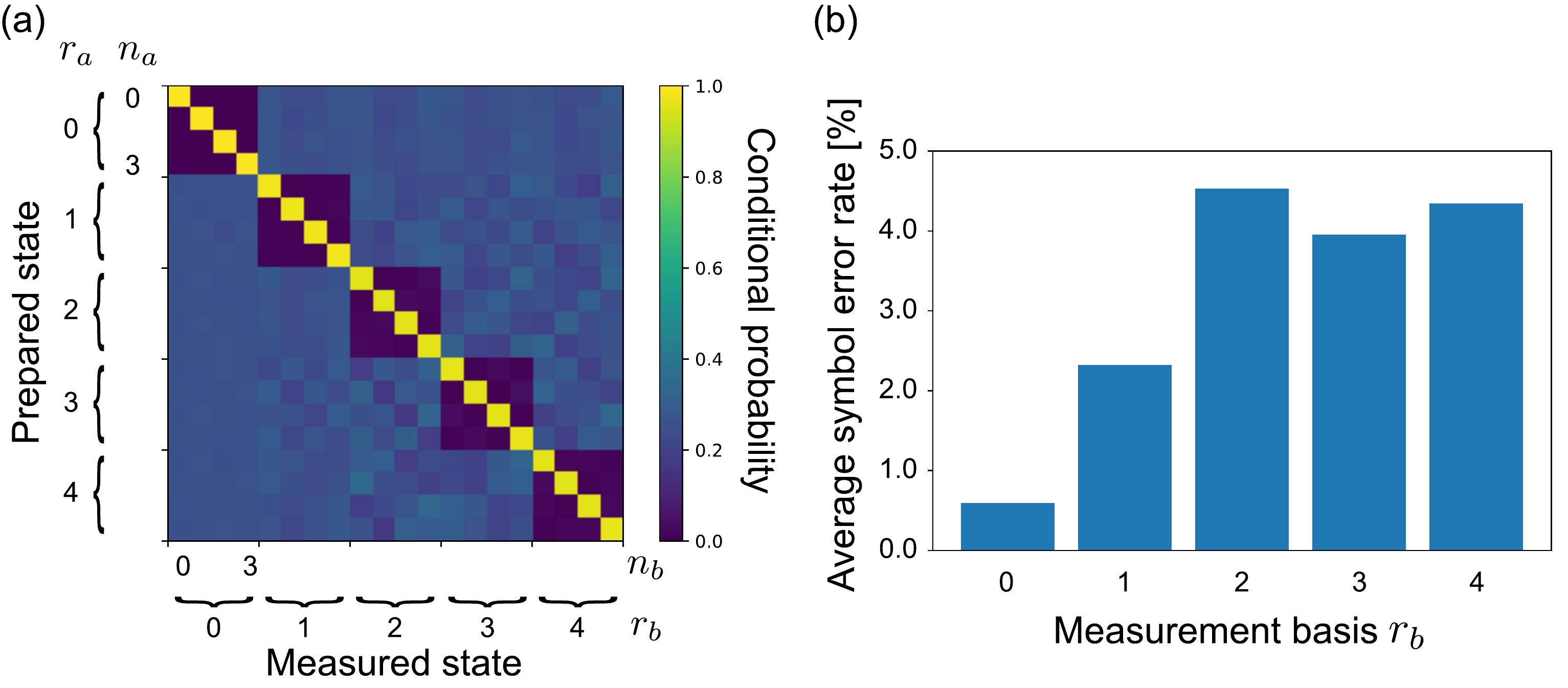}
    \caption{Experimental results.
    (a) Conditional probability of photon detection, ${\rm Pr}\left(n_b \middle| r_a, n_a, r_b \right)$.
    (b) Average symbol error rate for each basis estimated by $1 - \frac{1}{4} \sum_{n_a=n_b}{\rm Pr}\left(n_b \middle| r_a, n_a, r_b \right)$, where $r_a=r_b$.
    }
    \label{fig:results}
\end{figure}

Fig. \ref{fig:results}(a) shows a conditional probability distribution of photon detection ${\rm Pr}\left(n_b \middle| r_a, n_a, r_b \right)$,
where a photon was detected as the $n_b$th state
when it had been prepared as the $n_a$th state in the $r_a$th basis and measured by the $r_b$th basis.
Here, $r_a (r_b) = 0$ corresponds to the time-bin basis,
and $r_a (r_b) = r+1$ corresponds to the $r$th phase basis.
When the measurement basis was the same as the prepared basis,
the distribution was close to the identity matrix because each basis is an orthonormal basis.
On the other hand,
the distribution was close to the uniform distribution
when the state preparation and measurement were performed by different bases,
which formed MUBs.
Therefore,
we clearly observed the probability distribution expected from the five four-dimensional MUBs using the compact experimental setup.

Important parameters for QKD are error rates (Fig. \ref{fig:results}(b)),
which were estimated from the diagonal blocks of the conditional probability distribution.
In the time-bin basis ($r_b=0$),
the average error was the lowest thanks to the SNSPD,
which had a low timing jitter and dark count.
In the Hadamard basis ($r_b=1$),
the error rate increased slightly,
but it was clearly smaller than the error rates in the other phase bases.
The amplitude imbalance of the generated states and limited extinction ratio of the MZIs can be considered
as error sources for the Hadamard basis measurement.
Therefore,
the error could be mitigated by controlling the IQ modulation signal more precisely
and compensating for the relative transmittance between the short and long arms in the MZIs \cite{Lo2018,Lo2020,SupplementalMaterial}.
The difference in the error rates between the Hadamard basis and the other phase bases can be considered as the difference in operational conditions for the IQ modulator and PM.
As the Hadamard basis state is composed of $\ket{\pm}$ for the equivalent qubits,
the state has only real amplitudes,
and thus the IQ modulator operated as a simple amplitude modulator.
In addition,
the Hadamard basis state can be measured without driving the PM.
On the other hand,
the other phase bases have both real and imaginary amplitudes and require the basis selection by the PM.
Thus,
inaccurate bias voltages for the IQ modulator and distorted modulation signal for the PM would impose additional errors for these phase bases.
Therefore,
we can expect that the error rates for these phase bases can be mitigated
so that they are close to the error rate for the Hadamard basis with more precise controls of these parameters.

Unfortunately,
previous security analyses were limited to prime dimensions \cite{Sheridan2010,Ferenczi2012,Wang2021}.
However,
we can follow a similar analysis by using Weyl operators generalized by the Galois field \cite{SupplementalMaterial}.
Although the observed error rates depended on the basis,
all error rates were clearly smaller than the threshold value of 23.17\%,
below which we can generate secure keys in an asymptotic limit.
In addition,
the error rates were also comparable to or lower than the $\approx$ 4\% error rates using the Fourier basis \cite{Islam2017a}.
Important future work includes more thorough investigations,
e.g., a decoy-state method, finite key analysis, and various noise models,
to evaluate the secure key rate in a practical situation.
Note that the efficiency due to the basis mismatch is
not as small as $1/(d+1)$ in an asymmetric basis selection,
where the time-bin basis is selected to generate a raw key with a high probability,
while other $d$ phase bases are used to upper bound the amount of eavesdropped information.
The inefficiency due to the basis mismatch is negligible in an asymptotic limit,
although a finite key analysis should be included for an actual implementation.
In addition,
the $(d+1)$-basis protocol shows a strong robustness against correlated noise \cite{SupplementalMaterial},
and what kind of physical noise shows such a correlation is an interesting open question.
Although a practical implementation requires these investigations,
the present results indicate the feasibility of fast and robust secret key generation enabled by the $(d+1)$-basis protocol using the proposed setup.

Finally,
we explain the extension of the method to the case of the power of odd prime numbers.
When $d = p^N$ for odd prime number $p$,
$(d+1)$ MUBs can be constructed by the following equation instead of Eq. \eqref{eq:Bmnr} \cite{Wootters1989}.
\begin{equation}
    B_{mn}^{(r)} = \frac{1}{\sqrt{p^N}} \exp(\frac{2\pi}{p}i \left(\sum_{j=0}^{N-1} r_j \mathbf{m}^\mathrm{T} \mathbf{A}^{(j)} \mathbf{m} + \mathbf{m} \cdot \mathbf{n} \right)) . \label{eq:Bmnr_odd}
\end{equation}
Here, each element of $\mathbf{r}$, $\mathbf{m}$, $\mathbf{n}$, and $\mathbf{A}^{(j)}$ takes a value in $[0, \cdots, p-1]$,
and $\mathbf{A}^{(j)}$ is similarly constructed by using $GF[p^N]$.
In this case,
the phase takes only $p$ values,
which is a large reduction from $p^N$ values in the Fourier basis in Eq. \eqref{eq:fourie}.
Interestingly,
the minimum number of phases is obtained for $d=3^N$ because Eq. \eqref{eq:Bmnr} requires four phases.
This matrix can be also decomposed as $\mathbf{D}^{(r)}\mathbf{B}^{(0)}$,
and the diagonal unitary $\mathbf{D}^{(r)}$ can be similarly implemented by a phase modulator.
On the other hand,
$\mathbf{B}^{(0)}$ is not the Hadamard transform matrix
but a tensor product of the $p$-dimensional Fourier transform matrices.
Namely,
we need to project the state onto the tensor product of the $p$-dimensional Fourier basis states for the equivalent $p$-dimensional qudits.
Note that the tensor product states of the $p$-dimensional Fourier basis states are different from $p^N$-dimensional Fourier basis states for $N\geq 2$.
For a three-dimensional time-energy entanglement,
the Fourier basis measurement has been demonstrated by using a three-arm interferometer \cite{Thew2004},
which is a natural extension of $\ket{\pm}$ measurements by an  MZI.
Generally,
the $p$-dimensional Fourier basis measurement can be implemented by a $p$-arm interferometer.
Therefore,
the desired measurement corresponding to $\mathbf{B}^{(0)}$ can be implemented by cascading $p$-arm interferometers with different delays $N$ times via additional delay lines.

In conclusion,
we proposed a compact implementation of $(d+1)$ MUBs for prime power dimensions using $\log_p d$ interferometers.
The proposed method was demonstrated using a four-dimensional time-bin state,
where we observed the low average error rate of 3.8\% for the phase bases.
All error rates were below the threshold to distill a secure key against coherent attack,
and comparable to or lower than those in a previous implementation using the Fourier basis although the number of bases increased.
Our method can be also applied for other optical modes,
and constitutes an important step toward a practical implementation of fast, secure, and robust communications realized with a high-dimensional quantum state.

We thank K. Igarashi, K. Tamaki, and G. Kato for fruitful discussions and H. Tamura for administrative support in this research.

T.I. and S.A. contributed equally to this work.

\bibliography{mendeley}

\end{document}


\title{Supplemental Material: Scalable implementation of $(d+1)$ mutually unbiased bases\\for $d$-dimensional quantum key distribution}
\author{Takuya Ikuta}
\email[]{takuya.ikuta.xu@hco.ntt.co.jp}
\affiliation{NTT Basic Research Laboratories, NTT Corporation, 3-1, Morinosato Wakamiya, Atsugi, Kanagawa, 243-0198 Japan}
\affiliation{Graduate School of Engineering, Osaka University, 2-1, Yamadaoka, Suita, Osaka, 565-0871 Japan}

\author{Seiseki Akibue}
\email[]{seiseki.akibue.rb@hco.ntt.co.jp}
\affiliation{NTT Communication Science Laboratories, NTT Corporation, 3-1, Morinosato Wakamiya, Atsugi, Kanagawa, 243-0198 Japan}

\author{Yuya Yonezu}
\author{Toshimori Honjo}
\author{Hiroki Takesue}
\affiliation{NTT Basic Research Laboratories, NTT Corporation, 3-1, Morinosato Wakamiya, Atsugi, Kanagawa, 243-0198 Japan}

\author{Kyo Inoue}
\affiliation{Graduate School of Engineering, Osaka University, 2-1, Yamadaoka, Suita, Osaka, 565-0871 Japan}

\date{\today}

%
%
\renewcommand{\theequation}{S\arabic{equation}}
\renewcommand{\thesection}{S\arabic{section}}
\renewcommand{\thefigure}{S\arabic{figure}}

\maketitle

\section{Measurement setup for general optical modes}
In this section,
we explain how to extend the measurement of $(d+1)$ MUBs for general optical modes in detail.

First, let us summarize the case of time-bin states in the main text to extract their general functionalities.
To measure the states in the phase bases,
we need to implement two elements corresponding to $\mathbf{D}^{(r)}$ and $\mathbf{B}^{(0)}$.
$\mathbf{D}^{(r)}$ is implemented by a $\mathrm{LiNbO_3}$ phase modulator
because it can change the optical phase depending on time.
In other words,
we use a time-dependent phase modulator to implement $\mathbf{D}^{(r)}$ for time-bin states.
The projective measurement corresponding to $\mathbf{B}^{(0)}$,
the Hadamard basis,
is implemented by cascaded delay Mach-Zehnder interferometers (MZIs).
Each MZI has a different delay time corresponding to one of the equivalent qubits.
The functionality of a delay MZI is the time shift to make interference between different temporal modes.
Namely,
a delay MZI is used as an MZI having a temporal mode shift for time-bin states.
These MZIs are connected via optical delay lines.
The purpose of the delay lines is to avoid unnecessary interferences in the succeeding MZIs.
Although we use the temporal degree of freedom here,
this functionality is also available for general optical modes
as long as the delay times are carefully chosen to avoid unnecessary interferences between different inputs.
In addition,
other types of mode conversions can also be used instead of optical delay lines.
Finally,
a temporal filtering is performed for the measurement records to obtain the outcomes corresponding to the desired interferences in the central time slot.
The remaining Z basis measurement is performed by simply measuring the arrival time.
These functionalities for time-bin states are summarized as follows.
\begin{enumerate}
  \item $\mathbf{D}^{(r)}$ is implemented by a {\it time-dependent} phase modulator for {\it time-bin states}.
  \item The projection onto $\ket{\pm}$ for a equivalent qubit is performed by a MZI having a {\it temporal mode shift} for {\it time-bin states}.
  \item Optical delay lines are used to avoid unnecessary interferences,
  which are not limited to the temporal mode.
  \item {\it Temporal filtering} is used to retrieve the desired measurement results for {\it time-bin} states.
  \item Z basis measurement is performed by measuring the {\it arrival time} for {\it time-bin states}.
\end{enumerate}

\begin{figure}[tbp]
    \centering
    \includegraphics[width=18cm,pagebox=cropbox,clip]{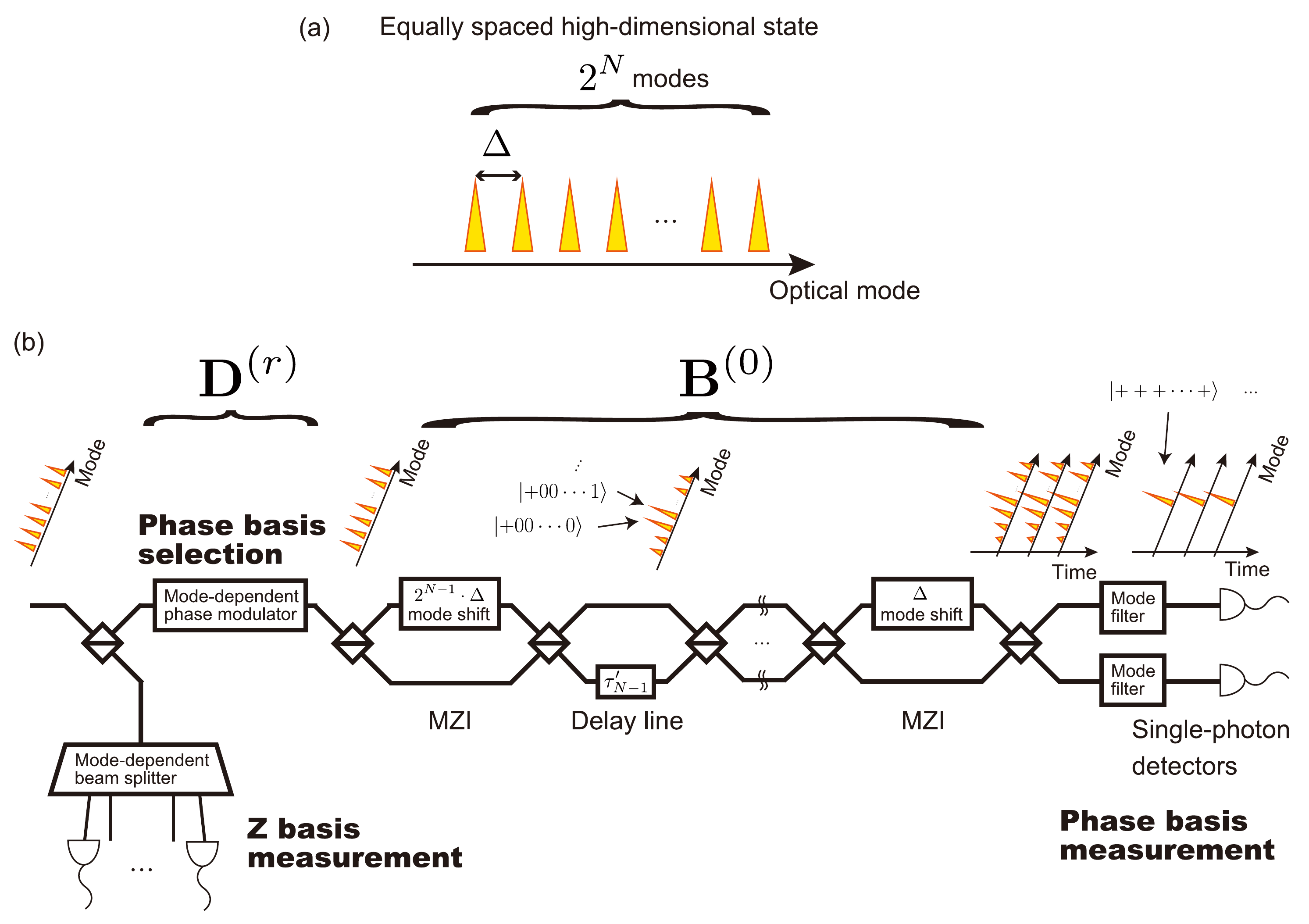}
    \caption{(a) $2^N$-dimensional quantum state in general optical mode.
    Each orthogonal mode is separated by $\Delta$.
    (b) Measurement setup of $(d+1)$ MUBs for general optical modes.}
    \label{fig:MUB_mes_for_general_modes}
\end{figure}

From these observations,
we can easily extend the proposed method for general optical modes.
Let us assume that a high-dimensional state is represented by $2^N$ equally spaced orthogonal states in the selected optical mode (Fig. \ref{fig:MUB_mes_for_general_modes}(a)).
Let $\Delta$ be the difference between the nearest neighbors in the selected optical mode.
The high-dimensional state is launched into a beam splitter,
where the Z basis and phase basis measurements are selected randomly (Fig. \ref{fig:MUB_mes_for_general_modes}(b)).
$\mathbf{D}^{(r)}$ is implemented by a mode-dependent phase modulator for a selected optical mode.
The modulated optical pulses are launched into cascaded MZIs.
The $k$th MZI ($k \in \{0, \cdots, N-1\}$) has a mode shift of $2^{N-k-1} \cdot \Delta$ in one arm.
The interference in the $k$th MZI corresponds to projective measurement onto $\ket{\pm}$ for the $(N-k-1)$th equivalent qubit.
These MZIs are connected via optical delay lines having delay times of $\tau_{N-k-1}^{'}$. The value of $\tau_{N-k-1}^{'}$ can be arbitrary chosen as long as unnecessary interferences are avoided.
Finally, the photon is detected by two sets of a mode filter and single-photon detector.
The measurement outcomes are recorded depending on the arrival time and ports.
For the Z basis measurement,
we can use a mode-dependent beam splitter followed by single-photon detectors.

A specific example is a high-dimensional frequency-bin state.
The Z basis measurement can be easily implemented by using a wavelength division multiplexing (WDM) filter.
A programmable filter can implement $\mathbf{D}^{(r)}$ for frequency-bin states \cite{Kues2017}.
In this case,
a delay MZI is replaced by a MZI having frequency shift for one arm.
A frequency shift can be implemented by a $\mathrm{LiNbO_3}$ optical modulator \cite{Kues2017,Lo2017} or by using optical nonlinearity \cite{Kumar1990,Takesue2010b}.
Optical delay lines are similarly used to avoid unnecessary interferences in succeeding MZIs.
Optical band-pass filters can be used as mode filters in front of the detectors,
where the filters remove the unnecessary results in the side peaks
corresponding to unnecessary time slots for time-bin states.
A frequency shift instead of delay is another choice to avoid unnecessary interferences,
especially for increasing amounts of information per unit time.
In this case,
we need to modify the detection to distinguish the measurement outcomes by the frequency shift instead of arrival times.
Although we need to use $d$ detectors,
additional WDM filters can be used instead of optical band-pass filters for this purpose.

Another example is an optical path mode,
with which a robust high-dimensional QKD using the Hadamard basis was recently demonstrated by using $d$-component interferometers with a tree structure \cite{Hu2021}.
As pointed out in \cite{Hu2020},
a spatial light modulator can be used to transform a phase basis to another phase basis.
By setting the phases according to Eq. (3) in the main text,
$(d+1)$ MUBs can be implemented.
In addition,
more compact interferometers can be constructed by using the cascaded structure shown in Fig. \ref{fig:MUB_mes_for_general_modes}(b) instead of the tree structure.

\section{Details of the experimental setup and calibration}

In this section,
we describe the details of the experiment shown in Fig. 2 in the main text.

A continuous-wave (CW) light, whose wavelength was 1559.0 nm,
was modulated into a single pulse or four sequential pulses by an optical in-phase and quadrature phase (IQ) modulator.
The pulse width, time interval, and repetition frequency of the state preparation were 33 ps, 500 ps, and 250 MHz, respectively.
These pulses were generated not randomly but sequentially because the purpose of the experiment was to evaluate the basic performance of the proposed method rather than to implement an actual QKD.
The optical power was attenuated by variable optical attenuators and optical couplers
to prepare weak coherent light.
The average photon number estimated by the optical power meter was $4.1 \times 10^{-3}$ photons per state.

In the phase basis measurement,
the weak optical pulses were launched into a phase modulator (PM)
to implement the phase basis selection $\mathbf{D}^{(r)}$.
The modulation pattern was also sequential.
The photon polarization was adjusted for the PM using a polarizer and polarization controller (PC).
Then,
the pulses modulated by the PM were launched into a stable MZI fabricated by a planar lightwave circuit (PLC) technology \cite{Honjo2004,Takesue2005}.
The MZI had a $\mathrm{SiO_2}$ core and clad deposited on a $\mathrm{Si}$ substrate.
The coupling loss between the PLC chip and optical fiber was 0.5 dB per coupling point,
and the excess loss for each beam splitter was 0.15 dB.
The 1-ns delay was introduced by the optical path 20.12 cm longer than the short path,
where the longer optical path resulted in additional optical loss of 0.8 dB.
These values of the MZI's characteristics are nominal values.
The relative phase between the two arms was adjusted to be 0
by stabilizing the chip temperature and tuning the voltage applied to a thermo-optic phase shifter on the long arm.
This calibration was initially performed using a CW light split from the original light source (not shown in the figure).
The extinction ratio of $>$20 dB was observed between the two output ports.
The MZI had a variable optical attenuator implemented in the short path
to compensate for the transmittance imbalance between the short and long arms \cite{Lo2018,Lo2020}.
Although we did not use the internal attenuator,
because the extinction ratio was high enough to perform the proof-of-principle experiment,
a higher extinction ratio can be obtained by using the attenuator. 
The pulses output from this MZI were launched into another MZI
after a relative delay of $\approx 250$ ps was introduced by two optical delay lines (DLs).
The delay time of the second MZI was 500 ps,
which corresponds to a 10.06-cm optical path and additional loss of 0.5 dB.
The relative phase was similarly adjusted to be 0.
The outputs from these cascaded MZIs were detected by using two superconducting nanowire single-photon detectors (SNSPDs).
The detection efficiencies were set at 56\%, and the dark counts were $<$ 100 cps.
The single count rates for SNSPD1 and 2 were 52 and 47 kcps, respectively.
These counts were recorded by a time-interval analyzer for 60 s.

When we performed the time-bin basis measurement,
the photons were directly detected by SNSPD1 after removing the PM and MZIs.
The single count rate was 546 kcps.
This measurement can be simultaneously performed by using a beam splitter when we implement an actual QKD protocol.
Due to the timing jitter of the detectors and electrical system,
we observed the full-width-at-half-maximum (FWHM) temporal width of 78 ps for the histogram of the arrival time records,
where the time-interval analyzer had a time resolution of 4 ps. 
In the data analysis,
we set the time window for each time slot at 200 ps,
which was larger than the FWHM temporal width for the histogram records. 

As described above,
the MZIs were calibrated by maximizing the extinction ratio between the output ports.
This is another practical advantage of the proposed method.
If we use the Fourier basis,
we need to use different calibrations for different MZIs because the relative phases take different values.
For the Hadamard basis,
the relative phases for all MZIs take only 0,
which simplifies the calibration procedure.

\section{Security analysis against coherent attack}

In this section,
we show that the asymptotic key rate derived in \cite{Sheridan2010} can also be used for the MUBs in our experiment.
To show such a generalization,
we use a different representation of MUBs constructed by the Galois field,
where the difference is only a permutation of the labels used in the main text.
In \ref{sec:Notation},
we describe basic rules of notations and arithmetic operations in the Galois field.
In \ref{sec:MUBs_for_qubits},
we show that the MUBs for $d=2^N$ introduced in \cite{DURT2010} are the same as the MUBs used in the main text.
In \ref{sec:MUBs_for_prime_powers},
we also show that the MUBs for odd prime power dimensions in \cite{Durt2005a,DURT2010}
are the same as the MUBs used in the main text,
although the equivalence itself was proven in \cite{Durt2005a}.
In \ref{sec:key_rate_against_collective},
we show the generalization of the key rate analysis in \cite{Sheridan2010}.
In \ref{sec:correlated_noise_model},
we analyze a correlated noise model
where the $(d+1)$-basis protocol has a larger improvement on the secret key rate
than an improvement for the depolarizing channel.

\subsection{Notation of MUB and arithmetic operations in the Galois field}
\label{sec:Notation}
We denote $r$th MUB in $\mathbb{C}^d$ by unitary operator
\begin{equation}
B^{(r)}:=(\ket{e^{(r)}_1},\cdots,\ket{e^{(r)}_d}),
\end{equation}
where
\begin{eqnarray}
\forall r, B^{(r)\dag}B^{(r)}=\mathbb{I},\\
\forall r\neq s,\forall i,j,|(B^{(r)\dag}B^{(s)})_{ij}|=\frac{1}{\sqrt{d}}
\end{eqnarray}
are satisfied.

We assume that $B^{(Z)}=\mathbb{I}$, i.e., the last MUB is the Z basis.

When $d=p^N$, where $p$ is a prime number,
it is known that there exists $(d+1)$ MUBs \cite{Wootters1989}, including the Z basis.
In the following, we suppose $d=p^N$ and the index of MUBs $r$ runs from $0$ to $d-1$. We often use a vector representation of the index such as $\vec{r}=(r_0,r_1,\cdots,r_{N-1})^T$, where $r=\sum_{n=0}^{N-1}r_np^n$. We often identify the natural number $r$ as an element of $GF[p^N]$ such that the addition $\oplus$ corresponds the elementwise addition mod $p$ in the vector representation. We also use set $\{A^{(k)}\}_{k=0}^{N-1}$ of $N$ by $N$ matrices related to the multiplication $\odot$ such that
\begin{equation}
p^m\odot p^n=\oplus_{k=0}^{N-1} A^{(k)}_{mn}p^k.
\end{equation}
Note that $A^{(k)}$ is symmetric, i.e., $A=A^T$, and invertible, i.e., $f_k:GF[p^N]\rightarrow GF[p^N]$ defined by $f_k(r)=A^{(k)}\vec{r} \mod p=\oplus_{m,n=0}^{N-1}A_{mn}^{(k)}r_np^m$ is bijective.

We also use $\ominus$ and $\oslash$ to denote the inverse operations of $\oplus$ and $\odot$, respectively.

%
%

\subsection{Equivalence of $N$-qubit MUBs}
\label{sec:MUBs_for_qubits}

Wootters and Fields \cite{Wootters1989} showed that for $r\in\{0,\dots,2^N-1\}$,
\begin{equation}
B^{(r)}_{ij}:=\frac{1}{\sqrt{2^N}}\exp\left(\frac{\pi}{2} i\left(\sum_{k=0}^{N-1} r_k\vec{i}^T A^{(k)}\vec{i}+2\vec{i}\cdot \vec{j}\right)\right)
\end{equation}
form MUBs.

On the other hand, Durt et al. \cite{DURT2010} claimed that 
\begin{equation}
 H^{(r)}_{ij}:=\frac{1}{\sqrt{2^N}}{\alpha_{\ominus i}^r}^*\gamma^{\ominus i\odot j} \label{eq:Durt_basis_states_even}
\end{equation}
form MUBs,
where $\gamma=-1$,
$\alpha_i^r:=\prod_{m,n=0}^{N-1}i^{r\odot (i_m2^m)\odot(i_n2^n)}$,
and $*$ denotes complex conjugate.

We show the equivalence between $B^{(r')}_{ij'}$ and $H^{(r)}_{ij}$ under a certain permutation. First,
\begin{eqnarray}
 \gamma^{\ominus i\odot j}&=& \gamma^{ i\odot j} \nonumber\\
 &=&\exp\left(\pi i\left(\vec{i}^TA^{(0)}\vec{j}\right)\right) \nonumber\\
 &=&\exp\left(\pi i\left(\vec{i}\cdot\overrightarrow{f_0(j)}\right)\right).
\end{eqnarray}
Second, by letting $r'=f_0(r)$, we obtain
\begin{eqnarray}
  &&{\alpha_{\ominus i}^r}^*={\alpha_i^r}^* \nonumber\\
&&=\prod_{m,n=0}^{N-1}(-i)^{r\odot (i_m2^m)\odot(i_n2^n)} \nonumber\\
&&=\prod_{m,n=0}^{N-1}(-1)^{\sum_{k,l=0}^{N-1}r_kA^{(1)}_{kl}i_mA^{(l)}_{mn}i_n} \nonumber\\
&&\times\prod_{m,n=0}^{N-1}(-i)^{\sum_{k=0}^{N-1}r'_ki_mA^{(k)}_{mn}i_n\mod 2}.  \label{eq:2nd_term_of_alpha}
\end{eqnarray}
By using the fact that $A^{(k)}$ is symmetric, the first term can be reduced to
\begin{eqnarray}
&& \prod_{m,n=0}^{N-1}(-1)^{\sum_{k,l=0}^{N-1}r_kA^{(1)}_{kl}i_mA^{(l)}_{mn}i_n} \nonumber\\
&&= \prod_{n=0}^{N-1}(-1)^{\sum_{k,l=0}^{N-1}r_kA^{(1)}_{kl}A^{(l)}_{nn}i_n^2} \nonumber\\
&&=\exp\left(\pi i\left(\vec{i}\cdot\vec{a}\right)\right),
\end{eqnarray}
where $a_n=\sum_{k,l=0}^{N-1}r_kA^{(1)}_{kl}A^{(l)}_{nn}\mod 2$.
Since $A^{(k)}$ is symmetric,
\begin{eqnarray}
  && \prod_{n\neq m}(-i)^{\sum_{k=0}^{N-1}r'_ki_mA^{(k)}_{mn}i_n\mod 2}  \nonumber\\
  && =\prod_{n \leq m}(-1)^{\sum_{k=0}^{N-1}r'_ki_mA^{(k)}_{mn}i_n\mod 2}  \nonumber\\
  && =\prod_{n \leq m}(-1)^{\sum_{k=0}^{N-1}r'_ki_mA^{(k)}_{mn}i_n}  \nonumber\\
  && =\prod_{n\neq m} i^{\sum_{k=0}^{N-1}r'_ki_mA^{(k)}_{mn}i_n} .
\end{eqnarray}

Then, the second term of ${\alpha_{\ominus i}^r}^*$ in Eq. \eqref{eq:2nd_term_of_alpha} can be reduced to
\begin{eqnarray}
 &&\prod_{m,n=0}^{N-1}(-i)^{\sum_{k=0}^{N-1}r'_ki_mA^{(k)}_{mn}i_n\mod 2} \nonumber\\
 &&=\prod_{n=0}^{N-1}(-i)^{\sum_{k=0}^{N-1}r'_kA^{(k)}_{nn}i_n\mod 2}\prod_{m\neq n}i^{\sum_{k=0}^{N-1}r'_ki_mA^{(k)}_{mn}i_n}\nonumber\\
 &&=\exp\left(\frac{\pi}{2} i\left(\sum_{k=0}^{N-1}r'_k\vec{i}^TA^{(k)}\vec{i}\right)\right)\nonumber\\
 &&\times\prod_{n=0}^{N-1}(-i)^{(\sum_{k=0}^{N-1}r'_kA^{(k)}_{nn}i_n\mod 2)+\sum_{k=0}^{N-1}r'_kA^{(k)}_{nn}i_n} \nonumber\\
  &&=\exp\left(\frac{\pi}{2} i\left(\sum_{k=0}^{N-1}r'_k\vec{i}^TA^{(k)}\vec{i}\right)\right)\exp\left(\pi i\left(\vec{i}\cdot\vec{b}\right)\right),
\end{eqnarray}
where $b_n=0$ if $\sum_{k=0}^{N-1}r'_kA^{(k)}_{nn}\in\{0,3\}\mod 4$ and $b_n=1$ if $\sum_{k=0}^{N-1}r'_kA^{(k)}_{nn}\in\{1,2\}\mod 4$. In summary, $H^{(r)}_{ij}=B^{(r')}_{ij'}$ if we use mappings $r'=f_0(r)$ and $j'=f_0(j)\oplus a\oplus b$. Since the mappings are bijective, $B^{(r')}_{ij'}$ and $H^{(r)}_{ij}$ are equivalent under a certain permutation.

\subsection{Equivalence of MUBs when $p\geq3$}
\label{sec:MUBs_for_prime_powers}

Wootters and Fields \cite{Wootters1989} showed that for $r\in\{0,\dots,2^N-1\}$,
\begin{equation}
B^{(r)}_{ij}:=\frac{1}{\sqrt{p^N}}\exp\left(\frac{2\pi}{p} i\left(\sum_{k=0}^{N-1} r_k\vec{i}^T A^{(k)}\vec{i}+\vec{i}\cdot \vec{j}\right)\right)
\end{equation}
form MUBs.

On the other hand, Durt et al. \cite{DURT2010,Durt2005a} claimed that 
\begin{equation}
 H^{(r)}_{ij}:=\frac{1}{\sqrt{p^N}}{\alpha_{\ominus i}^r}^*\gamma^{\ominus i\odot j}  \label{eq:Durt_basis_states_odd}
\end{equation}
form MUBs, where $\gamma=\exp\left(\frac{2\pi}{p} i\right)$ and $\alpha_i^r:=\gamma^{\ominus(r\odot i\odot i)\oslash 2}$.
We can easily show the equivalence between the two sets of MUBs as follows:
\begin{eqnarray}
{\alpha_{\ominus i}^r}^*&=&\gamma^{(r\oslash 2)\odot i\odot i} \nonumber\\
&=&\exp\left(\frac{2\pi}{p} i\sum_{k=0}^{N-1}f_0(r\oslash2)_k\vec{i}^TA^{(k)}\vec{i}\right)\\
 \gamma^{\ominus i\odot j}&=&\exp\left(\frac{2\pi}{p} i\left(\vec{i}\cdot \overrightarrow{f_0(\ominus j)}\right)\right).
\end{eqnarray}
Therefore, $H^{(r)}_{ij}=B^{(r')}_{ij'}$ if we use mappings $r'=f_0(r\oslash 2)$ and $j'=f_0(\ominus j)$. Since the mappings are bijective, $B^{(r')}_{ij'}$ and $H^{(r)}_{ij}$ are equivalent under a certain permutation.

\subsection{Error rate threshold against coherent  attack}
\label{sec:key_rate_against_collective}

Because the secure key rate in \cite{Sheridan2010,Ferenczi2012} was evaluated by using Bell basis states generalized by Weyl operators, $U_{ij}$,
the analysis for the $(d+1)$-basis protocol is valid only for prime dimensions.
That means, strictly speaking,
the analysis cannot be applied for prime power dimensions in our case.
However, the analysis in \cite{Sheridan2010} can be extended for prime power dimensions by using MUBs expressed by Eq. \eqref{eq:Durt_basis_states_even} or \eqref{eq:Durt_basis_states_odd}
and by Weyl operators generalized by the Galois field, $V_{ij}$ \cite{DURT2010}.
Here, we focus on the security analysis against collective attack because of its simplicity.
However,
it is known that the asymptotic secret key rate evaluated for collective attack is equal to the one evaluated for coherent attack
by using random permutation of qudits and quantum de Finetti theorem \cite{Renner2005,Renner2007}.
In this work,
we do not consider a finite key length effect;
thus, the following security analysis is also valid for coherent attack.

The two types of Weyl operators are similarly defined as follows.
\begin{eqnarray}
  U_{ij} &&= \sum_{k=0}^{d-1} \omega^{kj} \ket{k+i \!\!\!\mod d} \bra{k}, \\
  V_{ij} &&= \sum_{k=0}^{d-1} \gamma^{(k\oplus i)\odot j} \ket{k \oplus i} \bra{k},   \label{eq:Generalized_Weyl_Operator}
\end{eqnarray}
where $\omega$ is the $d$th root of unity,
and $i,j \in \{0, 1, \cdots, d-1\}$.
Note that the basis state $\ket{e_k^{(r)}}$ defined by Eq. \eqref{eq:Durt_basis_states_even} or \eqref{eq:Durt_basis_states_odd} is the eigenstate of $V_{l, r\odot l}$ for $^\forall l$.
In addition,
$V_{ij}$ satisfies
\begin{eqnarray}
V_{ij}\ket{e_k^{(r)}} &&= \gamma^{i \odot k} {\alpha_i^r}^* \ket{e_{r \odot i \ominus j \oplus k}^{(r)}}, \label{eq:state_shift_on_rbasis}\\
V_{ij} \ket{e_k^{(Z)}} &&= \gamma^{(k\oplus i)\odot j} \ket{e_{i \oplus k}^{(Z)}}. \label{eq:state_shift_on_Zbasis}
\end{eqnarray}
Let $\ket{\Phi_{00}} = \frac{1}{\sqrt{d}} \sum_{s} \ket{ss}$
and $\ket{\Phi_{ij}} = \left(V_{ij} \otimes \mathbb{I}\right) \ket{\Phi_{00}}$, which are generalized Bell basis states.
With these prerequisites,
the analysis in \cite{Sheridan2010} can be extended for prime power dimensions.

First, Alice prepares $\ket{\Phi_{00}}$ and sends one of the qudits to Bob.
Alice randomly selects the basis $r$ and measures the remaining qudit by $\{\ket{e_a^{(r)}}\}$.
Bob performs a similar measurement using $\{\ket{{e_b^{(r')}}^*}\}$, where $*$ denotes the complex conjugate of the probability amplitudes expanded by Z basis states.
From the measurement outcomes when Alice and Bob use the same basis $r$,
they estimate the following error vectors.
\begin{equation}
  \underline{q}_r = \{q_r^0, q_r^1, \cdots, q_r^{(d-1)}\}, \label{eq:error_vector}
\end{equation}
where $q_r^t = {\rm Pr}\left(a \ominus b = t \middle| r \right)$,
and $a$ and $b$ are measurement outcomes for Alice and Bob.
$\underline{q}_Z$ is similarly defined for the Z basis.
From Eq. \eqref{eq:state_shift_on_rbasis} and \eqref{eq:state_shift_on_Zbasis},
both measurements $\{\ket{e_a^{(r)} {e_b^{(r)}}^*}\}$ and $\{ V_{ij} \otimes V_{ij}^* \ket{e_a^{(r)} {e_b^{(r)}}^*} \}$ for $a\ominus b=t$ correspond to the same error.
Thus,
$\underline{q}_r$ does not change if Alice and Bob simultaneously apply $V_{ij}^\dag$ and $V_{ij}^T$ on their qudits before the measurement.
Therefore,
we can estimate $q_r^t$ using the state $\widetilde{\rho}_{AB}$ shared between Alice and Bob as
\begin{eqnarray}
q_r^t && = \sum_{a\ominus b = t}
  \bra{e_a^{(r)} {e_b^{(r)}}^*}
   \widetilde{\rho}_{AB}
  \ket{e_a^{(r)} {e_b^{(r)}}^*}  \nonumber\\
&&= \sum_{a\ominus b = t}
  \bra{e_a^{(r)} {e_b^{(r)}}^*}
    \frac{1}{d^2} \sum_{ij}
      V_{ij}^\dag \otimes V_{ij}^T
        \widetilde{\rho}_{AB}
      V_{ij} \otimes V_{ij}^*
  \ket{e_a^{(r)} {e_b^{(r)}}^*}. \label{eq:VV_sandwitched}
\end{eqnarray}
The generalized Bell basis state satisfies $V_{ij}\otimes V_{ij}^* \ket{\Phi_{i'j'}}= \gamma^{i'\odot j \ominus i \odot j'} \ket{\Phi_{i'j'}}$;
thus,
\begin{eqnarray}
&&\bra{\Phi_{i'j'}}
  \sum_{ij}
    V_{ij}^\dag \otimes V_{ij}^T
      \widetilde{\rho}_{AB}
    V_{ij} \otimes V_{ij}^*
\ket{ \Phi_{ i'' j'' }}  \nonumber\\
&&=
\sum_{ij}
  \bra{\Phi_{i'j'}}
    \gamma^{\ominus i'\odot j \oplus i \odot j'}
      \widetilde{\rho}_{AB}
    \gamma^{i'' \odot j \ominus i \odot j''}
  \ket{\Phi_{i''j''}}  \nonumber\\
&&=
\sum_{ij}
  \gamma^{i \odot (j' \ominus j'') \ominus (i' \ominus i'') \odot j }
  \bra{\Phi_{i'j'}}
      \widetilde{\rho}_{AB}
  \ket{\Phi_{i''j''}}  \nonumber\\
&& = 
d^2 \delta_{i', i''} \delta_{j', j''}
\bra{\Phi_{i'j'}}
  \widetilde{\rho}_{AB}
\ket{\Phi_{i''j''}}.
\end{eqnarray}
This equation indicates that
the state in Eq. \eqref{eq:VV_sandwitched} is diagonalized in the generalized Bell basis states.
Therefore,
from the same reasoning as in \cite{Sheridan2010,Kraus2005a},
we can treat the state shared between Alice and Bob
as if it has the following form.
\begin{equation}
  \rho_{AB} = \sum_{j,k=0}^{d-1} \lambda_{jk} \ket{\Phi_{jk}} \bra{\Phi_{jk}}, \label{eq:Bell_diagonal}
\end{equation}
where $\lambda_{jk} \geq 0$ and $\sum_{jk} \lambda_{jk} = 1$.
From Eq. \eqref{eq:Bell_diagonal},
the error vector is expressed by
\begin{eqnarray}
  q_Z^t &&= \sum_{k=0}^{d-1} \lambda_{t k}, \label{eq:evaluate_q_Z_by_lambda}\\
  q_k^t &&= \sum_{j=0}^{d-1} \lambda_{j, k\odot j \ominus t} \quad\text{for}\quad k\in \{0, \cdots, d-1\}. \label{eq:evaluate_q_r_by_lambda}
\end{eqnarray}
These equations indicate that
\begin{equation}
  \lambda_{jk} = \frac{1}{d} \left( \sum_{s=0}^{d-1} q_s^{s\odot j \ominus k} + q_Z^j -1\right). \label{eq:lambda_jk}
\end{equation}

Equation  \eqref{eq:error_vector} and Eqs. \eqref{eq:Bell_diagonal}--\eqref{eq:lambda_jk} correspond to Eqs. (1)--(4) in \cite{Sheridan2010}.
By using these generalizations for prime power dimensions,
the remaining analysis in \cite{Sheridan2010} holds.
Therefore,
the threshold value of the error rate can be evaluated by the same equations in \cite{Sheridan2010} as follows.
\begin{eqnarray}
  \chi(A:E) &&= H_{d^2}(\underline{\lambda}) - H_{d}(\underline{q}_Z), \\
  I(A:B) &&= \log_2 d - H_{d}(\underline{q}_Z), \\
  r_{\infty} &&= I(A:B) - \chi(A:E)  \nonumber\\
    &&= \log_2 d - H_{d^2}(\underline{\lambda}). \label{eq:asymp_rate_detailed}
\end{eqnarray}
Here,
$\chi(A:E)$ is Eve's information upper bounded by the Holevo bound,
$I(A:B)$ is the mutual information between Alice and Bob's outcomes,
$r_{\infty}$ is the asymptotic secrete key rate,
$H_{d'}(\underline{p})$ is the $d'$-dimensional entropy function for a probability distribution $\underline{p}$,
and $\underline{\lambda}$ is a vector whose element is $\lambda_{jk}$.
In brief,
the analysis in \cite{Sheridan2010} can be applied for prime power dimensions simply with replacements of equations using $\!\!\!\mod d$ by arithmetic operations in the Galois field.
Because the Galois field of order $d=p$ is isomorphic to an integer set $\{0, \cdots, d-1\}$
using arithmetic operations associated with $\!\!\!\mod d$,
this is a natural generalization for prime power dimensions.

\begin{table}[htb]
 \begin{center}
  \caption{$\lambda_{jk}$ evaluated from the experimental results.}
  \begin{tabular}{crrrr}
    & \multicolumn{4}{c}{$k$}\\ \cline{2-5}
    $j$ &0&1&2&3 \\ \cline{1-5}
    0 & 0.9606 &  0.0084 &  0.0159 &  0.0091 \\
    1 & 0.0052 & -0.0039 &  0.0049 & -0.0040 \\
    2 & 0.0047 &  0.0046 & -0.0053 & -0.0022 \\
    3 & 0.0063 & -0.0029 & -0.0047 &  0.0033 \\ \cline{1-5}
  \end{tabular}
  \label{tab:lambda}
 \end{center}
\end{table}

We also tried to evaluate the asymptotic secret key rate using these equations and the experimental results.
However,
the estimation of $\lambda_{jk}$ failed because some of the results took small negative values, which were not physical
(Table \ref{tab:lambda}).
Similar problems are often observed in quantum state tomography for a high-fidelity state,
where such problems are usually circumvented by the maximally likelihood estimation \cite{James2001}.
Although a more detailed analysis is required for a provably secure and practical QKD,
we can derive a reference value of $r_{\infty}$ to evaluate the potential of the proposed method,
by using the upper bound of $H_{d^2}(\underline{\lambda})$.
\begin{eqnarray}
  H_{d^2}(\underline{\lambda}) &&= \sum_{j,k} -\lambda_{jk} \log_2 \lambda_{jk}  \nonumber\\
    &&= -\lambda_{00} \log_2 \lambda_{00} - \sum_{(j,k) \neq (0,0)} \lambda_{jk} \log_2 \lambda_{jk}  \nonumber\\
    &&= -\lambda_{00} \log_2 \lambda_{00}
         - (1-\lambda_{00}) \log_2 (1-\lambda_{00})
         + (1-\lambda_{00}) \sum_{(j,k) \neq (0,0)} -\frac{\lambda_{jk}}{1-\lambda_{00}} \log_2 \frac{\lambda_{jk}}{1-\lambda_{00}}  \nonumber\\
    &&\leq -\lambda_{00} \log_2 \lambda_{00}
         - (1-\lambda_{00}) \log_2 (1-\lambda_{00})
         + (1-\lambda_{00}) \log_2 (d^2 - 1)  \nonumber\\
    &&= -\lambda_{00} \log_2 \lambda_{00}
         - (1-\lambda_{00}) \log_2 \frac{1-\lambda_{00}}{d^2 - 1} . \label{eq:upper_bound_H_d^2}
\end{eqnarray}
The inequality follows from the fact that the maximal value of $H_{(d^2-1)}(\underline{p})$ is $\log_2 (d^2-1)$.
From Eq. \eqref{eq:lambda_jk},
$\lambda_{00}$ is given by
\begin{eqnarray}
  \lambda_{00} &&= \frac{1}{d} \left( \sum_{r=0}^{d-1} q_r^{0} + q_Z^0 -1\right)  \nonumber\\
    &&= \frac{1}{d} \left(d - \sum_{r=0}^{d-1} e_r - e_Z \right)  \nonumber\\
    &&= 1 - \frac{d+1}{d}\overline{e} , \label{eq:lambda00_by_err_avg}
\end{eqnarray}
where we define $e_r$ and $e_Z$ as symbol error rates for $r$ and Z bases, respectively,
and $\overline{e}$ as the symbol error rate averaged over all bases.
Eqs. \eqref{eq:asymp_rate_detailed}, \eqref{eq:upper_bound_H_d^2} and \eqref{eq:lambda00_by_err_avg}
give the lower bound of the secret key rate.
In prime dimensions,
this key rate is equal to the previous results obtained under the assumption of a depolarizing channel \cite{Sheridan2010} or several symmetries,
including the probability of basis selection \cite{Ferenczi2012}.
Note that we did not assume any specific channel model or additional symmetries;
thus,
the secret key rate formula using the average symbol error rate gives a conservative lower bound against any coherent attack with asymmetric basis selection.

From the experimental results,
the lower bound of the secret key rate was evaluated as $r_{\infty} \geq 1.6$ 
with $\lambda_{00} = 0.96$.
Therefore,
the experimental results implied the potential secret key generation over 1.6 bits per photon count,
which cannot be achieved by the BB84 or six-state protocol even without errors.
However,
because the negative $\lambda_{jk}$ could indicate some imperfections in the implementations as well as finite length effects \cite{Wang2021},
we should keep this value as just a reference value,
and it is important in the future to develop a more elaborate method which bounds the secret key more conservatively and practically.

\subsection{Strongly correlated noise model}
\label{sec:correlated_noise_model}

As shown in Eq. \eqref{eq:asymp_rate_detailed},
the secret key rate for the $(d+1)$-basis protocol is evaluated by $\lambda_{jk}$,
which is the coefficient of the generalized Bell basis states in Eq \eqref{eq:Bell_diagonal}.
Because $\ket{\Phi_{ij}} = \left(V_{ij} \otimes \mathbb{I}\right) \ket{\Phi_{00}}$,
the labels $i, j$ of $\lambda_{ij}$ can be interpreted as labels of errors induced by $V_{ij}$.
Here,
the generalized Weyl operator $V_{ij}$ can be decomposed by two generalized Weyl operators as follows \cite{DURT2010}.
\begin{equation}
  V_{ij} = V_{0j}V_{i0} .   \label{eq:decomposition_of_generalized_Weyl}
\end{equation}
From Eq. \eqref{eq:Generalized_Weyl_Operator}
\begin{align}
  V_{i0} &= \sum_{k=0}^{d-1} \ket{k\oplus i}\bra{k}  \nonumber\\
    &= \sum_{k=0}^{d-1} \ket{e_{k\oplus i}^{(Z)}}\bra{e_{k}^{(Z)}}  .
\end{align}
The above equation indicates that
$V_{i0}$ shifts the measurement outcomes for the Z basis by $i$ in the Galois field.
Therefore,
$V_{i0}$ is an $i$ shift operator in the Z basis.
Similarly,
$V_{0j}$ can be interpreted as a $\ominus j$ shift operator in the $0$th phase basis because
\begin{align}
  V_{0j} &= \sum_{k=0}^{d-1} \gamma^{k\odot j}\ket{k}\bra{k} 
    = \sum_{k=0}^{d-1} \ket{e_{k}^{(0)}}\bra{e_{k \oplus j}^{(0)}}   \nonumber\\
    &= \sum_{k=0}^{d-1} \ket{e_{k \ominus j}^{(0)}}\bra{e_{k}^{(0)}}  .
\end{align}
From these observations,
$\lambda_{ij}$ can be considered as
a joint probability distribution of $i$ and $\ominus j$ shift error in the Z and phase bases.
In the two-basis protocol,
we can obtain only the marginal probability distributions of the two types errors,
$\underline{q}_Z$ and $\underline{q}_0$.
In this sense,
the basic difference between the two-basis protocol and $(d+1)$-basis protocol
is the information about correlation between two types of errors.
Therefore,
the $(d+1)$-basis protocol can have a large information gain
when we consider a correlated noise as pointed out for the six-state protocol \cite{Koashi2009}.

To illustrate the effect of correlated errors,
let us consider strongly correlated shift errors in the Z and phase bases,
where only $\lambda_{ii}$ has non-zero value
and $\lambda_{ij}$ is zero for $i \neq j$.
In addition, we assume $\lambda_{ii}$ for $i \neq 0$ are equal for the simplicity.
From Eqs. \eqref{eq:evaluate_q_Z_by_lambda} and \eqref{eq:evaluate_q_r_by_lambda},
the error vectors in the Z basis and 0th phase basis are given by
\begin{align}
  q_Z^t &= \sum_{k=0}^{d-1} \lambda_{tk} = \lambda_{tt} ,   \\
  q_0^t &= \sum_{j=0}^{d-1} \lambda_{j, \ominus t} = \lambda_{\ominus t, \ominus t} .
\end{align}
Therefore, $\lambda_{00} = 1-e_Z$
and $\lambda_{ii} = e_Z/(d-1)$ for $i \neq 0$.
An example of $\lambda_{ij}$ with this model is shown in Table \ref{tab:lambda_of_correlated_noise}.
These equations indicate a strong correlation between two errors
where a $t$ shift error in the Z basis
coincides with a $\ominus t$ shift error in the 0th phase basis.
\begin{table}[htb]
 \begin{center}
  \caption{$\lambda_{jk}$ for correlated noise model for $d=4$.}
  \begin{tabular}{ccccc}
    & \multicolumn{4}{c}{$k$}\\ \cline{2-5}
    $j$ &0&1&2&3 \\ \cline{1-5}
    0 & $1 - e_Z$ &  0&  0&  0\\
    1 & 0& $e_Z/3$&  0& 0\\
    2 & 0&  0& $e_Z/3$ & 0\\
    3 & 0& 0& 0& $e_Z/3$ \\ \cline{1-5}
  \end{tabular}
  \label{tab:lambda_of_correlated_noise}
 \end{center}
\end{table}

From Eq. \eqref{eq:evaluate_q_r_by_lambda},
the error rates for other phase bases can be derived as follows.
\begin{align}
  e_r = 1-q_r^0 = 1 - \sum_{j=0}^{d-1} \lambda_{j, r\odot j} =
  \begin{cases}
    e_Z & \text{for} \quad r \neq 1 \\
    0 & \text{for} \quad r = 1
  \end{cases} .   \label{eq:phase_symbol_errors_for_correlated_noise}
\end{align}
Here, we used $r \odot j = j$ if and only if $r=1$ or $j=0$.
Therefore,
the phase basis for $r=1$ has no error,
while the other phase bases have the same error as the Z basis.
This implies that no privacy amplification is required for the $(d+1)$-basis protocol against this correlated noise,
while it is required for the two-basis protocol.
From Eq. \eqref{eq:phase_symbol_errors_for_correlated_noise},
the average symbol error rates for the two protocols are given by
\begin{equation}
  \bar{e} = \begin{cases}
    e_Z & \text{for the two-basis protocol} \\
    \frac{d}{d+1} e_Z & \text{for the $(d+1)$-basis protocol}
  \end{cases} .
\end{equation}

By using the secret key rate formula for the two-basis protocol \cite{Sheridan2010},
the secrete key rate for the correlated noise model is given by
\begin{align}
  r_{\text{two-basis}}^{\text{cor}} &= \log_2 d - H_{d}(\underline{q}_Z) - H_{d}(\underline{q}_0)   \nonumber\\
    &= \log_2 d + 2 \left\{
      (1-e_Z) \log_2(1-e_Z)
      +e_Z \log_2 \left(\frac{e_Z}{d-1}\right)
    \right\}     \nonumber\\
    &= \log_2 d + 2 \left\{
      (1-\bar{e}) \log_2(1-\bar{e})
      +\bar{e} \log_2 \left(\frac{\bar{e}}{d-1}\right)
    \right\}  .
\end{align}
Note that this secret key rate is also equal to the rate derived for a depolarizing channel.
On the other hand,
the secret key rate for the $(d+1)$-basis protocol is given by
\begin{align}
  r_{\text{$(d+1)$-basis}}^{\text{cor}} &= \log_2 d - H_{d^2}(\lambda)   \nonumber\\
    &= \log_2 d + (1-e_Z) \log_2(1-e_Z)
      +e_Z \log_2 \left(\frac{e_Z}{d-1}\right)   \nonumber\\
    &= \log_2 d + \left(1-\frac{d+1}{d}\bar{e}\right) \log_2\left(1-\frac{d+1}{d}\bar{e}\right)
      + \frac{d+1}{d}\bar{e} \log_2 \left\{\frac{d+1}{d(d-1)}\bar{e}\right\}  .
\end{align}

\begin{figure}[tbp]
    \centering
    \includegraphics[width=10cm,pagebox=cropbox,clip]{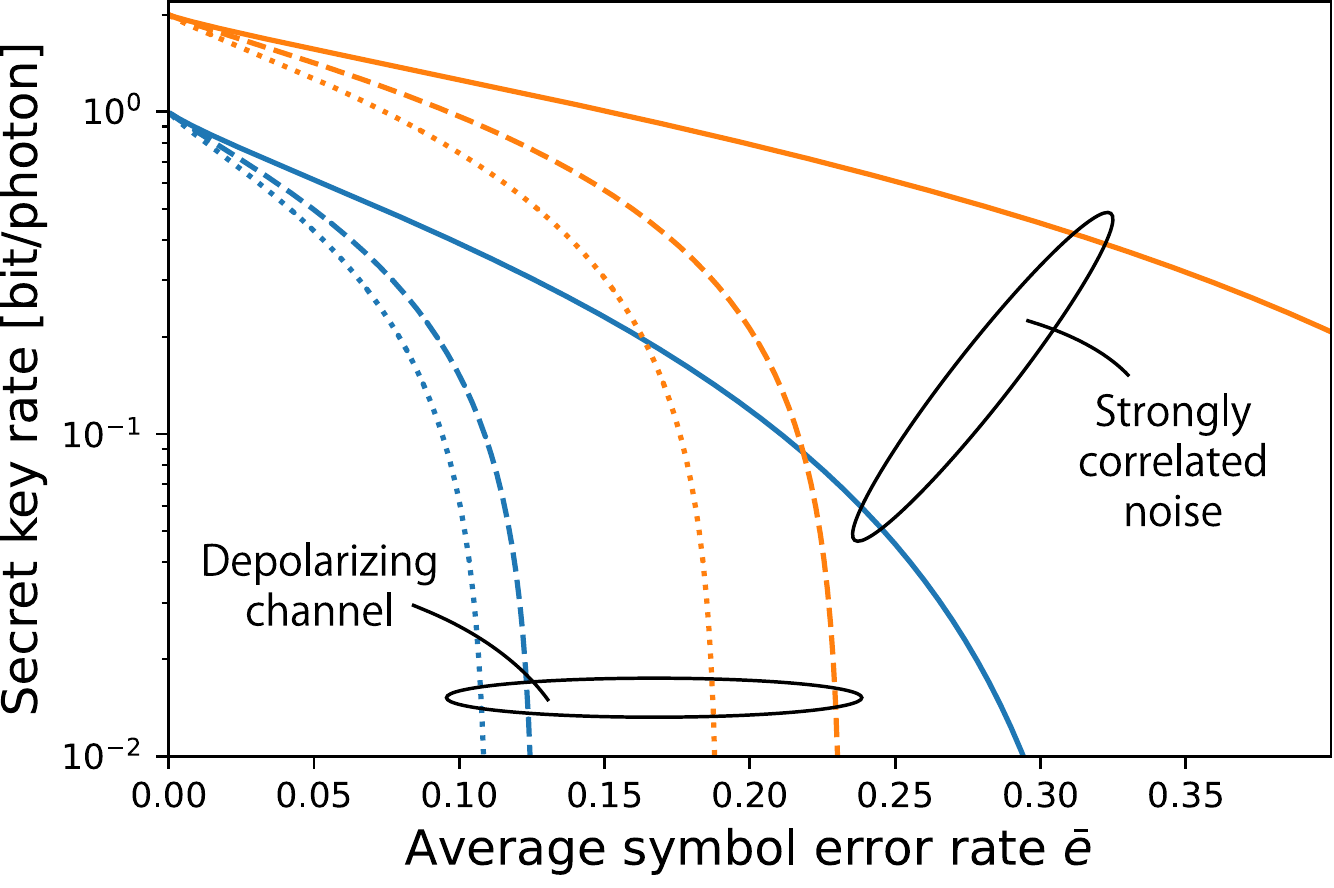}
    \caption{Secret key rate depending on average symbol error rate.
    The blue and orange lines show the results for $d=2$ and $4$, respectively.
    The solid, dashed and dotted lines correspond to
    the $(d+1)$-basis protocol with a strongly correlated noise,
    the $(d+1)$-basis protocol with a depolarizing channel,
    and the two-basis protocol, respectively.}
    \label{fig:secret_key_rate_for_correlated_noise}
\end{figure}

Fig. \ref{fig:secret_key_rate_for_correlated_noise} shows the secret key rate
depending on the noise models, protocols, and the number of dimensions.
The two-basis protocol gives the same secret key rate independent of the two noise models.
In the $(d+1)$-basis protocol,
the secret key rate increased for the depolarizing channel slightly
while it increased largely for the strongly correlated noise.
These results imply that
the information about the correlation between two types of errors can largely improve the secret key rate in the $(d+1)$-basis protocol.

Although the above noise model can illustrate an advantage of the $(d+1)$-basis protocol,
more careful and thorough investigations are required for a realistic situation
because the above noise model is an extreme example.
As shown in Eq. \eqref{eq:phase_symbol_errors_for_correlated_noise},
the phase basis for $r=1$ has no error.
This means that,
if we use the phase basis for $r=1$ instead of $r=0$ in the two-basis protocol,
the two-basis protocol achieves an identical secret key rate for the same $e_Z$ in the $(d+1)$-basis protocol.
However,
such an optimization of the basis requires a prior knowledge of the optical channel.
Therefore,
we need to construct a different optical system depending on the optical channel,
which would vary its characteristics depending on a time in a worse case.
On the other hand,
the $(d+1)$-basis protocol can obtain the information about correlation of errors for a general situation.
Thus,
we can expect to obtain a larger chance where the improved key rate is achieved.
In a realistic situation,
the secret key rate of the $(d+1)$-basis protocol should be plotted
between the solid and dashed lines in Fig. \ref{fig:secret_key_rate_for_correlated_noise}
because the secret key rate for the depolarizing channel gives the lower bound for the same average symbol error rate
as described in \ref{sec:key_rate_against_collective}.
Note that we need to solve a problem about negative $\lambda_{jk}$ to employ the full advantage using the correlation of errors.
Although more concrete examples should be investigated as a thorough comparison to demonstrate a practical advantage in the future,
we believe $(d+1)$ bases are beneficial for enhancing the robustness of the high-dimensional QKD against general correlated errors.

\bibliography{mendeley_for_SI}

